\documentclass[%
 prl,%
 amsmath,amssymb,
 reprint,%
superscriptaddress
]{revtex4-2}

\usepackage{graphicx}% Include figure files
\usepackage{dcolumn}% Align table columns on decimal point
\usepackage{bm}% bold math
\usepackage{xcolor}
\usepackage[normalem]{ulem}
\usepackage[english]{babel}
\usepackage{float}

\usepackage{physics}
\usepackage{textcomp}
\usepackage{siunitx}

\usepackage{verbatim} 
\usepackage{color}
\usepackage{subfigure} 
\usepackage{hyperref}

\usepackage{xr}
\usepackage{nameref}

\usepackage{ifthen}
\usepackage{amssymb}
\usepackage{xcolor}

\newboolean{showcomments}
\setboolean{showcomments}{true}

\makeatletter
\newcommand{\mynote}[3]{%
  \ifthenelse{\boolean{showcomments}}{%
   \fbox{\bfseries\sffamily\scriptsize#1}%
   {\small$\blacktriangleright$\textsf{\emph{\color{#3}{#2}}}$\blacktriangleleft$}}%
  {%
   \@bsphack
   \@esphack
  }%
}
\makeatother

\definecolor{asparagus}{rgb}{0.53, 0.66, 0.42}
\begin{document}
%\preprint{AIP/123-QED}

\title{A superconducting quantum circuit single artificial atom maser}

\author{Maria Mucci}
\affiliation{Department of Physics and Astronomy, University of Pittsburgh, Pittsburgh, PA, 15260, USA}

\author{Nicholas Hougland}
\affiliation{Department of Physics and Astronomy, University of Pittsburgh, Pittsburgh, PA, 15260, USA}
\affiliation{Pittsburgh Quantum Institute, University of Pittsburgh, Pittsburgh, PA, 15260, USA}

\author{Chun-Che Wang}
\affiliation{Department of Applied Physics, Yale University, New Haven, CT, 06511 , USA}

\author{Israa Yusuf}
\affiliation{Department of Physics and Astronomy, University of Pittsburgh, Pittsburgh, PA, 15260, USA}

\author{Chenxu Liu}
\affiliation{Pacific Northwest National Laboratory, Richland, WA 99352, USA}

\author{David Pekker}
\email[email: ]{pekkerd@pitt.edu}
\affiliation{Department of Physics and Astronomy, University of Pittsburgh, Pittsburgh, PA, 15260, USA}
\affiliation{Pittsburgh Quantum Institute, University of Pittsburgh, Pittsburgh, PA, 15260, USA}

\author{Michael Hatridge}
\email[email: ]{michael.hatridge@yale.edu}
\affiliation{Department of Physics and Astronomy, University of Pittsburgh, Pittsburgh, PA, 15260, USA}
\affiliation{Department of Applied Physics, Yale University, New Haven, CT, 06511 , USA}

\date{\today}% It is always \today, today, but any date may be explicitly specified

\begin{abstract}
We demonstrate a circuit QED analog of an atomic micromaser that utilizes an artificial, multi-level atom, pumped into a population-inverted state by a microwave tone, as the gain medium. Our demonstration is enabled by the flexibility of the circuit QED platform, which allowed us to precisely engineer the level-structure, coupling, and dissipation of the micromaser components. Our device shows rich physics and perhaps points to ways to use the recent developments in the domain of microwave quantum circuits to probe the domain of maser physics.
\end{abstract}

\maketitle

{\it Introduction -- \/}
\label{sec:Introduction}
%%% For PRL introduction should be no more than 2 paragraphs, let's try to make this shorter ...
Lasers and masers are ubiquitous in physics and engineering as stable, ultra narrow sources of optical and microwave frequency light~\cite{NewMaser_Gordon1955, QuantumOptics_Scully1997}. A maser consists of a gain medium that is coupled to a cavity which itself is coupled to an output port. To power the maser, the gain medium is pumped into a population inverted state, the excitations from the gain medium are then transferred to the cavity via stimulated emission, and the photons from the cavity are emitted into the output port. A key feature of the maser is that it can convert incoherent pump light (or current) into much more coherent output light. In conventional masers, the tunability of the properties of the gain medium is limited by the fact that it is composed of three- or four-level atoms or molecules. Circuit Quantum Electro-Dynamics (cQED)~\cite{cQED_Blais2004} offers a flexible alternative, in which the properties of the gain medium can be continuously tuned by varying the  superconducting quantum circuit layout to engineer an artificial atom and its interaction with the maser cavity.

Past work, both experimental~\cite{SingleAtomLasing_Astafiev12007, SingleCPLaser_Chen2014} and theoretical~\cite{PersistentLaser_You2007,VoltBiasLaser_Ashhab2009}, has focused on using current to pump a charge qubit into a population inverted state. The qubit was coupled to a microwave cavity and the maser was found to operate when the frequency of the qubit matched an integer multiple of the cavity frequency. A related scheme, that was described theoretically~\cite{JJLaserTheory_Simon2018} and observed experimentally~\cite{acJJLaser_Cassidy2017}, relied on the ac Josephson effect to directly pump a microwave cavity.

In this paper, we report on the operation and theoretical modeling of the cQED equivalent of an atomic micromaser (a maser in which the gain medium is a single atom). Specifically, we utilize a multi-level artificial atom as the gain medium which we drive into a population inverted state using a microwave pump. The engineering of the energy level-structure, matrix-element-structure, and dissipation of the artificial atom~\cite{ChemPot_Mucci2025} is enabled by constructing it from two nonlinear modes: a lossy SNAIL~\cite{SNAIL_Frattini2017} resonator with strong three-wave mixing coupled to a long-lived transmon~\cite{Transmon_Koch2007} qubit. This allows us to investigate the rich physics associated with the population dynamics of the artificial atom and its interplay with the maser cavity. Moreover, our experimental setup offers a great deal of in-situ tunability: in addition to being able to adjust the parametric pump that drives population inversion (like conventional atomic optics experiments) we can also adjust the transition frequencies and the nonlinearity of our artificial atom without having to fabricate a new device (or switching atom species). 

Our main results are as follows. First, using experimental observations and theoretical modeling we determine that our maser can function using three different pump cycles, in which the cavity is pumped by either the $\ket{e} \rightarrow \ket{g}$ or $\ket{f} \rightarrow \ket{e}$ one-photon transition of the transmon, or the $\ket{f} \rightarrow \ket{g}$ two-photon transition of the transmon. Second, we observe that our maser emits coherent light, locked to the cavity frequency, over a $\sim 50 \, \text{MHz}$ range of pump frequencies, which is consistent with our model of the maser. Third, using frequency-domain measurements we observe that the linewidth of output maser light can be as narrow as $54$~Hz (as compared to the $19.7$~kHz bare cavity linewidth). This observation is supported by our real-time measurement of the diffusion of the maser phase in the IQ plane.

% A pair of theoretical investigation, Refs.~\onlinecite{HeisenbergMaser_Baker2021, Maser_Liu2021} showed that microwave circuits could be designed to 

% Other works have also theorized and built unique masers out of Josephson junction-based circuits. The theory of cQED maser systems are featured in Refs.~\onlinecite{VoltBiasLaser_Ashhab2009} and \onlinecite{PersistentLaser_You2007}. Experimental results are found in Refs.~\onlinecite{SingleAtomLasing_Astafiev12007} and \onlinecite{SingleCPLaser_Chen2014}, which uses a charge qubits as artificial atoms and require voltage biasing to inject photons into the cavity. Use of the ac Josephson effect is theorized in Ref.~\onlinecite{JJLaserTheory_Simon2018} and experimentally realized in Ref.~\onlinecite{acJJLaser_Cassidy2017} with injection locking. Surpassing the standard quantum limit on laser linewidth is discussed in Ref.~\onlinecite{BeyondSTL_Wiseman1999} and proposed with the addition of one nonlinear inter-element coupler in Ref.~\onlinecite{HeisenbergMaser_Baker2021}. The maser presented in this work is based on Ref.~\onlinecite{Maser_Liu2021} which proposes the use of two nonlinear couplers between the atom and cavity and the cavity and its output to surpass the standard quantum limit by a factor of $\frac{1}{\langle {n} \rangle ^2}$ for photon number $n$. We first prove that linear couplings using the proposed artificial atom and cavity makes a maser with far decreased linewidth compared to the bare cavity.

\begin{figure}
    \centering
    \includegraphics[width = 0.97\columnwidth]{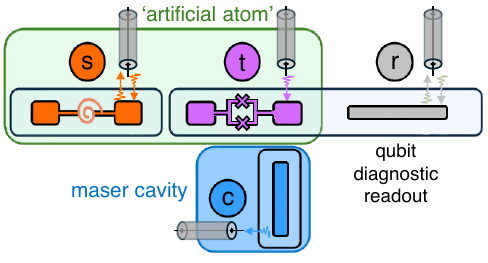}
    \caption{Schematic of the micromaser circuit. The circuit is composed of three parts: an artificial atom (green), a maser cavity (blue), and a qubit diagnostic readout cavity (gray). The artificial atom integrates a SNAIL (orange) coupled to a flux-tunable transmon qubit (purple), enabling parametric control and nonlinearity. Each component has its own transmission line, allowing independent pumping and measurement. The direction of the arrows indicates the direction of the RF signals used in this experiment. \label{fig:fig1}}
\end{figure}

\begin{figure*}
    \centering
    \includegraphics[width = 1\linewidth]{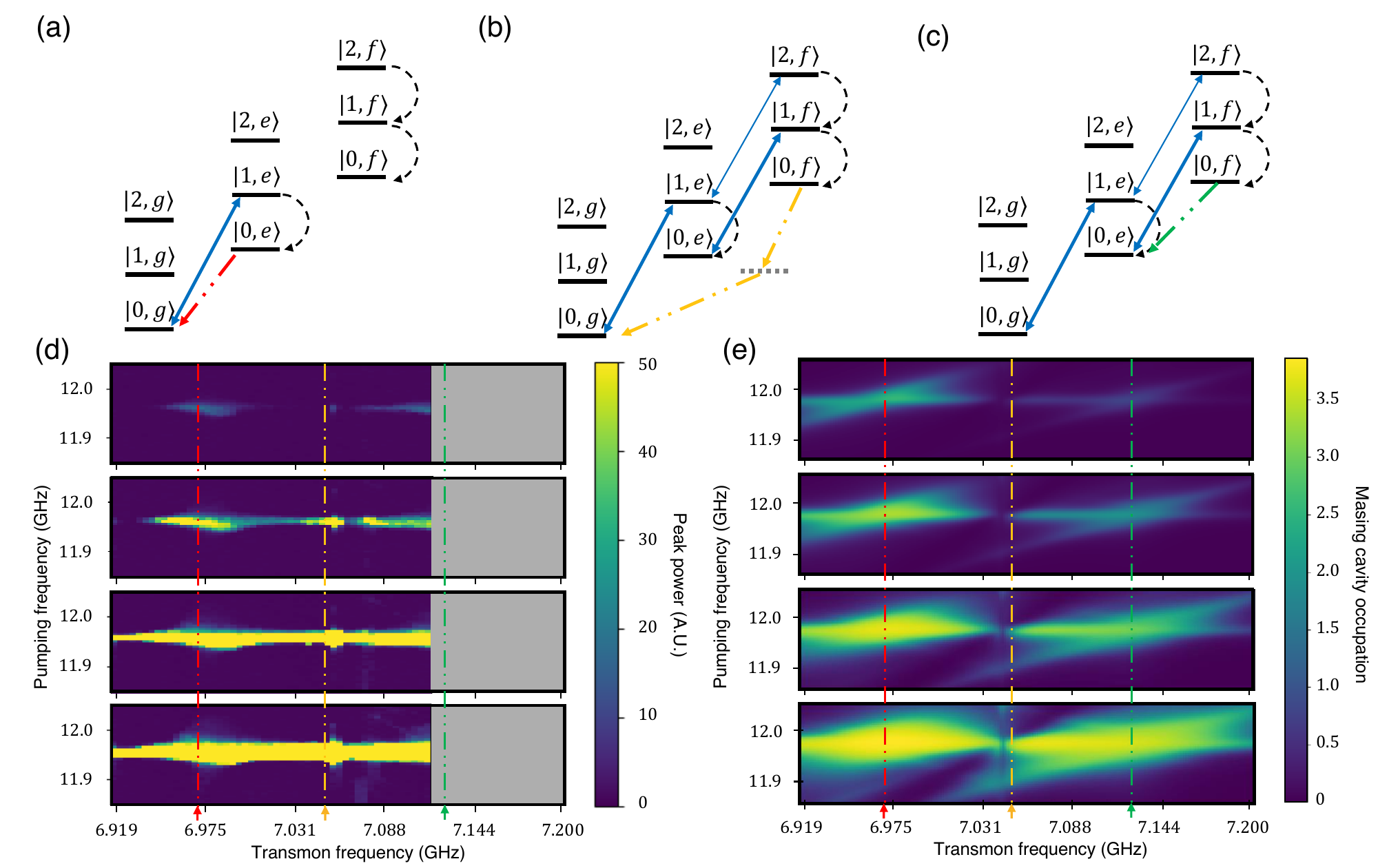}
    \caption{\label{fig:maserLevels}Diagram of levels in the SNAIL-transmon subsystem. Blue arrows represent couplings due to the parametric pump, black dashed arrows show relaxation of the SNAIL, and dot-dashed arrows indicate atom-cavity coupling. (a) Masing mediated by the transition between the transmon's first excited state and ground state. (b) Masing mediated by the transition between the transmon's second excited state and ground state. (c) Masing mediated by the transition between the transmon's second excited state and first excited state. (d) Maximum peak power from signal analyzer measurement from cavity port for varying transmon and pump frequencies. From the top to bottom panels, the pump power and qubit inversion rate $\Gamma^p_{ge}/2\pi$ are $P_0$: $0.178$~MHz, $2 \times P_0$: $0.475$~MHz, $4 \times P_0$: $1.19$~MHz, and $8 \times P_0$: $2.81$~MHz. 
    %Pumping DAC values are 10.6, 15, 21.2, and 30 from the top to bottom panel %respectively, which is doubled from top to bottom panel in pumping power 
    (e) Occupation of the masing cavity according to our simulation for varying transmon and pump frequencies. We show maser brightness for increasing pump power from the top to the bottom panel. We associate the bright regions at the lowest, middle, and highest transmon frequency with the ordinary $ge$ masing scheme (a, red), alternative $gf/2$ masing scheme (b, yellow), and alternative $ef$ masing scheme (c, green)}
    % We associate the bright region at the lowest transmon frequency with the ordinary masing scheme (a, red), while the bright region at the higher transmon frequency represents the first alternative masing scheme (b, yellow), and the bright region at the highest transmon frequency represents the second alternative masing scheme (c, green). \add{better numbers and label g2eff so this is not just totally wibley}} 
     
\end{figure*}

{\it The micromaser circuit -- \/}
\label{sec:Circuit description}
%There are many physical forms in which a maser may be realized. This experiment uses superconducting quantum circuit elements, letting us use qubits and parametric mixing elements comprised of Josephson junctions to invert the atom's population and build the photon number in a superconducting cavity. 
The schematic of our micromaser circuit is depicted in Fig.~\ref{fig:fig1}. The circuit consists of an artificial atom, a maser cavity, and an ancillary readout cavity. The artificial atom is composed of a SNAIL coupled to a transmon qubit; it is pumped through a transmission line that is coupled to the SNAIL resonator. Photons are transferred from the artificial atom to the maser cavity via a weak coupling between the transmon and the cavity. The maser light is output through a transmission line that is weakly coupled to the maser cavity. For diagnostic purposes, the circuit also has a readout cavity that is coupled to the transmon and a third transmission line. The readout cavity is only used to dispersively probe the properties of the artificial atom~\cite{DispersiveReadout_Wallraff2005}, bypassing the maser cavity. During normal operation, all transitions of the artificial atom are far detuned from the readout cavity frequency.

The SNAIL and maser cavity are fabricated on separate sapphire chips while the transmon and the readout cavity share a sapphire chip. The sapphire chips are housed within a T-shaped arrangement of tunnels~\cite{Tube_Axline2016} inside a 6061 aluminum enclosure with copper inserted for magnetic biasing. The two cavities are $\lambda /2$ tantalum strip-line resonators, where the geometry of the superconductor determines the frequency and coupling to other elements and external ports. The transmon features two nominally identical Josephson junctions in parallel which allow us to tune its frequency with externally applied magnetic field. This tunability is used to bring the transmon into resonance with the cavity. The SNAIL has one small junction in parallel with two larger junctions. The large features of the SNAIL and transmon are made of tantalum~\cite{Tantalum_Place2021} and the junctions are made of Al/AlOx/Al. Dissipation engineering is accomplished by coupling transmission lines and circuit elements via strategically placed microwave coupling pins and adjusting the SNAIL nonlinearity by magnetic field biasing. 

%, Our atom is `artificial,' meaning its mode structure is completely controlled by chosen circuit parameters. Our `atom' is a frequency-tunable transmon qubit that is weakly hybridized to a SNAIL coupler. Our masing cavity is a standard superconducting resonator mode $m$ with a weak dipole-dipole coupling to the transmon. 

{\it Micromaser operation and pump cycles -- \/}
\label{sec:data_section}
In order to illustrate the maser design choices and its operation, we will now qualitatively describe the processes involved in the one-photon $|e\rangle \rightarrow |g\rangle$ pump cycle, which is depicted schematically in Fig.~\ref{fig:maserLevels}a. To describe the quantum state of the artificial atom we use the $\ket{\text{SNAIL}, \text{transmon}}$ Fock basis, with $0, 1, 2, ...$ enumerating the number of photons in the SNAIL mode and $g, e, f, ...$ the number of photons in the transmon (e.g. the state $\ket{0, e}$ has no photons in the SNAIL and one photon in the transmon).

% \begin{table}
%     \caption[\textbf{Step-wise masing with parametric process and engineered loss}]{\textbf{Step-wise masing with parametric process and engineered loss} The masing process may be broken into individual steps that are all happening simultaneously, each through their own exchange, loss, or parametric pump, and each with its own rate. The empty qubit $\ket{0, g}$ is pumped at rate $g_{\Sigma_{s, ge}}$ and the SNAIL quickly decays at rate $\kappa_s$, creating the inverted transmon state $\ket{0, e}$. The qubit's photon is exchanged into the cavity at rate $g_{tc}$, putting a photon in the cavity. The process repeats to achieve a cavity population $\ket{N}$ at the combined 'upward' rate.}
%     \begin{tabular}{c | c | c | c }
%     \textbf{Initial state} & \textbf{Process} & \textbf{Target state} & \textbf{Rate}\\ \hline
%     $\ket{0, g, n}$ & $\Sigma_{s, ge}$ & $\ket{1, e, n}$ & $g_{\Sigma_{s, ge}}$ \\ \hline
%     $\ket{1, e, n}$ & SNAIL loss &  $\ket{0, e, n}$ & $\kappa_s$ \\ \hline
%     $\ket{0, e, n}$ & qubit-cavity exchange & $\ket{0, g, n + 1}$ & $g_{tc}$ \\ \hline
%     $\ket{0, g, n + 1}$ & Repeat above & $\ket{0, g, N}$ & $\Gamma_{ge}^p + g_{tc}$ \\ \hline
%     \end{tabular}
%     \label{tab:Maser_ladder}
% \end{table}

In our maser, the SNAIL acts as a parametric third order mixing element~\cite{PhasePreserveAmp_Bergeal2010, Paramp_Roy2016}. By pumping the SNAIL at the frequency $\omega_p$ that is roughly the sum frequency of the SNAIL and transmon modes $\omega_p\approx \omega_s + \omega_t$, we can induce a process in which a single pump photon is converted into one photon in the SNAIL qubit and one photon in the transmon qubit. This parametric process drives the $|0,g\rangle \rightarrow |1,e\rangle$ transition in the artificial atom (blue arrow in Fig.~\ref{fig:maserLevels}a). We have engineered the dissipation of the artificial atom in such a way that the the loss rate of the SNAIL, $\kappa_s$, is much higher than that of the transmon qubit. As a result the SNAIL quickly decays to its ground state, leaving the artificial atom in the population-inverted state $\ket{0,e}$ (dashed black arrow in Fig.~\ref{fig:maserLevels}a). We can quantify this incoherent process of bringing the artificial atom to the $\ket{0,e}$ by an effective up rate $\Gamma_{ge}^{p}$. As long as $\Gamma_{ge}^{p}$ is larger than the transmon decay rate, the artificial atom will maintain population inversion~\cite{ChemPot_Mucci2025}. 

In order for the masing process to take place, the transmon is biased to its flux condition that puts it on resonance with the maser cavity. This allows the photon to swap from the transmon into the cavity at a rate dictated by the transmon-cavity coupling matrix element $g_{tc}$. Following the swap, the cavity gains one more photon and the artificial atom returns to the ground state $\ket{0,g}$ (red arrow in Fig.~\ref{fig:maserLevels}a) before being re-excited to the $\ket{0,e}$ state by the parametric drive/engineered loss.  Masing occurs when the rate at which photons are added to the maser cavity exceed the bare loss rate of the maser cavity $\Gamma_c$, which is accompanied by the decrease of the maser cavity linewidth $\Gamma_c^{masing}$ well below its bare value.  In summary, we have the following hierarchy of scales:
\begin{equation}
\kappa_s \gg \Gamma_{ge}^{p} > g_{tc} \gg \Gamma_c. \label{eq:hierarchy}
\end{equation}
The SNAIL loss rate is larger than the effective up-rate of the artificial atom, which is larger than the coupling of the artificial atom to the maser cavity, which itself is larger than the cavity loss rate.

The complex level structure of our artificial atom enables richer patterns of dynamics than the conventional 3-level atom maser dynamics that we have described thus far. Specifically, we have observed two additional pump cycles that involve higher transmon states. The first alternative scheme involves masing via the $\ket{0, f} \rightarrow \ket{0, e}$ transition that is activated by first modifying the pump to achieve population inversion in the $\ket{0, f}$ state and then by making the $\ket{0, f} \rightarrow \ket{0, e}$ transition resonant with the level spacing of the cavity, as shown by Fig.~\ref{fig:maserLevels}c. The second alternative scheme activates a two-photon process~\cite{TwoPhotonMaser_Brune1987, TwoPhotonReview_Gauthier2003} associated with the $\ket{0, f} \rightarrow \ket{0, g}$ transition of the artificial atom shown in Fig.~\ref{fig:maserLevels}b. To activate this transition, the transition frequency $( \omega_{\ket{0, f}} - \omega_{\ket{0, g}}) /2$ must be resonant with the cavity's transition frequency. 

\begin{figure}
    \centering
    \includegraphics[width = 1.0\linewidth]{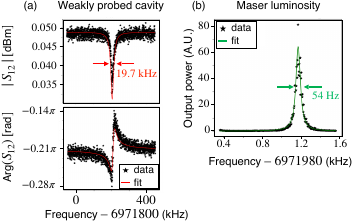}
    \caption{ Comparison of the (a) weakly probed `bare' maser cavity and (b) maser light with the maser tuned for narrowest linewidth and brightest luminosity. This translates to a nearly $365$ times decrease in cavity linewidth due to the masing process.}
    \label{fig:Lowest_BW}
\end{figure}  

\begin{figure*}
    \centering
    \includegraphics[width = \linewidth]{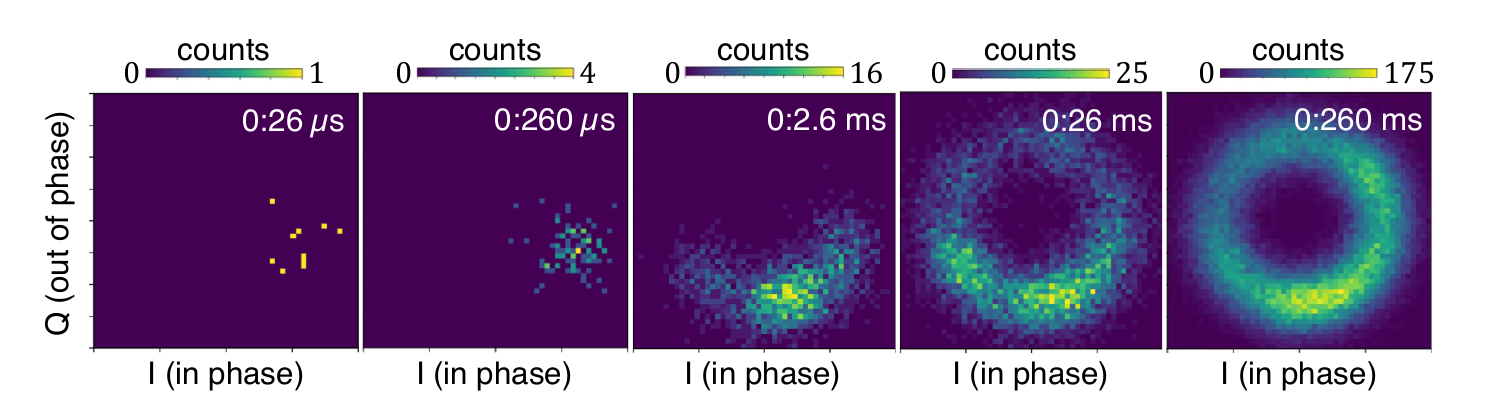}
    \caption{Time-domain measurement of the maser output-light. The in- and out-of-phase quadratures were measured every $2.6\,\text{ns}$ and integrated every $1000$ clock cycles. The five panels of the figure show histograms of data collected over different time-spans (as indicated).   
    % The output of the maser cavity can be connected to a digitizing channel and its signal can be demodulated into in-phase and quadrature components. Following the same procedure as shown in Fig~\ref{fig:SAsweep_narrowest}, we capture the maser's output and map to I and Q every $2.6$ ns. Here, we show cumulative measurements over the indicated time scales. First, we show the first 0 to $26~\mu$s of measurement. The maser appears in a small section of the IQ plane. The first $260~\mu$s show a more concentrated region of results. The maser is still coherent. As time continues, the maser spreads in I and Q, forming a ring. When a full ring has formed, we can no longer clearly distinguish the path of the maser in time or it it no longer coherent.
    }
    \label{fig:TimeRings_narrowest}
\end{figure*}

\makeatletter
\phantomsection % create a hyperlink target
\def\@currentlabelname{Micromaser operation and pump cycles} % set name for \nameref
\label{sec:data_section}
\makeatother

{\it Micromaser operation and pump cycles -- \/}
To observe the various pump cycles, we connect the maser cavity's output to a spectrum analyzer and measure the maser luminosity as a function of the parametric pump frequency and the transmon frequency for various settings of the parametric pump amplitude (see Fig.~\ref{fig:maserLevels}d). We use room temperature current sources to flux bias the SNAIL and transmon. The SNAIL mode is tuned to a fixed frequency at a point of high third- and low fourth-order nonlinearity~\cite{KerrFreeSNAIL_Sivak2019} while the transmon qubit is tuned through a range of frequencies in the vicinity of the cavity frequency. The pump frequency is tuned through a range of frequencies in the vicinity of the SNAIL and transmon modes' sum frequency. The pump tone is applied as a box car signal from a QICK RFSoC 111 DAC channel~\cite{QICK_Stefanazzi2022} mixed with the LO of a signal generator at room temperature, sent through the transmission line coupled to the SNAIL. The amplitude of this drive varies with the chosen voltage output of the DAC. The details of the tune-up process are provided in the online supplement.

When the pump amplitude is set to the lowest two settings (top two panels of Fig.~\ref{fig:maserLevels}d), the maser luminosity  has three distinct peaks as a function of the transmon frequency. These peaks correspond to three different resonant conditions between the transmon qubit and the cavity: $\omega_e-\omega_g=\omega_c$ (indicated by the red arrow), $\left(\omega_f-\omega_g\right)/2=\omega_c$ (indicated by the yellow arrow), and $\omega_f-\omega_e=\omega_c$ (indicated by the green arrow, this peak is partially cutoff in the experimental data). We have established the assignment of these resonances by weakly probing the maser cavity and observing avoided crossings as the transmon is tuned through resonance with the cavity (see online supplement). At higher pump amplitudes the three peaks broaden and merge into a single luminous band. We attribute the three peaks to the three pump cycles depicted in Fig.~\ref{fig:maserLevels}a-c. 

{\it Theoretical modeling of the pump cycles -- \/}
We validate the above attribution by performing a master equation simulation of our maser system (see the online supplement for additional details). When performing the numerical simulations we set the values of all parameters to our best measurement/estimate of their experimental values except for the pump amplitude. As we do not have an accurate measurement of the attenuation of the transmission line used to pump the artificial atom at low temperature, when performing simulations we choose the pump amplitude so as to best reproduce the experimental observations. Similarly, we do not have an accurate calibration for the maser luminosity and therefore for the output of the simulation we plot the occupancy of the maser cavity as opposed to an absolute luminosity. Due to the computational complexity of the numerical simulations, we cut off the photon number in the maser cavity at a maximum of four photons. Therefore, the photon number should be treated as a lower bound that is easily saturated at higher pump amplitudes.

Our numerical simulations show three distinct peaks (see Fig.~\ref{fig:maserLevels}e) and confirm that the peaks are indeed induced by the three pump cycles. The location and shape of the $\omega_e-\omega_g=\omega_c$ and $\omega_f-\omega_e=\omega_c$ peaks matches experimental observations quite well. On the other hand, the $\left(\omega_f-\omega_g\right)/2=\omega_c$ peak is displaced towards higher pump frequencies and is much dimmer in simulation than in the experiment, being visible only for the two highest pump amplitudes. We do not have a good explanation for this discrepancy. 

%We comment that a maser is a system that should be able to turn incoherent pump light into highly coherent output light. While we have not tested this directly, we do have strong circumstantial evidence that our maser is capable of doing just that. Specifically, both experimental observations and numerical testing show that our maser is relatively insensitive to the pump frequency; the maser remains luminous when the pump frequency is varied over $\approx 20-50\,\text{MHz}$ which is much wider than the bare maser cavity linewidth $\Gamma_c = 6-20$~kHz (the variation is due to transmon qubit hybridization). The insensitivity of our maser to the frequency of the pump light is engineered into the SNAIL qubit, which has a linewidth of $\kappa_s \approx 20-30$~MHz, and is therefore able to drive population inversion over a broad range of pump frequencies. 

{\it Observed lower bound on the maser linewidth -- \/} We use a spectrum analyzer to monitor the maser output light and adjusting the SNAIL frequency, transmon frequency, pump amplitude, and pump frequency until we find an operating point that produces a spectral peak that is both ultra narrow and bright. In Fig.~\ref{fig:Lowest_BW} we compare the characteristics of the bare maser cavity (a) to the maser light at the optimal operating point (b). We observe that the linewidth of the maser light is $54~\text{Hz}$, almost $365$ times narrower than the cavity's bare linewidth of $19.7$~kHz. 

In order to validate the frequency-domain measurement of the maser linewidth, we perform a time domain measurement. We monitor the phase drift of maser output light, by first connecting the maser cavity output to the spectrum analyzer and find its peak output frequency, we then immediately switch its output to go towards the ADC on the QICK FPGA where we record the integrated in-phase and out-of-phase quadrature every $2.6~\mu$s. We observe that the output light phase performs a random walk around the around the IQ plane, shown in  Fig.~\ref{fig:TimeRings_narrowest}. From the histograms, we estimate that the phase correlation time, or  the time it takes the phase to diffuse by $2\pi$, is about $26$~ms. The maser linewidth is the inverse of the phase correlation time. The time-domain estimate of the maser linewidth is $38$~Hz, which is consistent with the $54$~Hz linewidth that we obtained using frequency-domain measurement.

{\it Discussion -- \/}  We have theoretically analyzed, fabricated, and experimentally probed a cQED analog of an atomic micromaser -- one that utilizes a multi-level artificial atom that is pumped by microwave light.
Our device allowed us to probe a rich set of maser physics, including frequency- and time- domain measurement of phase diffusion, observation of the narrowing of the maser linewidth, and powering the maser with three distinct pumping cycles. We imagine that the control and flexibility that the cQED platform has brought to quantum computers can also be harnessed to explore maser physics.  As an example, the cQED platform could potentially be used to engineer a maser that surpasses the standard quantum limit of Schawlow and Townes on the maser linewidth~\cite{Maser_SchawlowTownes1958}, as theoretically described in Refs.~\cite{BeyondSTL_Wiseman1999, Maser_Liu2021,HeisenbergMaser_Baker2021}.

{\it Acknowledgements -- } 
We wish to acknowledge helpful discussions and support from Gurudev Dutt, Mingkang Xia, Guarav Agarwal, and Chao Zhou.  The TWPA used for amplified readout in this experiment was provided by MIT Lincoln Laboratory. This research was sponsored by M. Hatridge’s NSF CAREER grant (PHY-1847025) and the Army Research Office (under Award Numbers W911NF15-1-0397, W911NF-18-1-0144 [HiPS], and W911NF-23-1-0252 [FastCARS]). The views and conclusions contained in this document are those of the authors and should not be interpreted as representing the official policies, either expressed or implied, of the Army Research Office or the U.S. Government. The U.S. Government is authorized to reproduce and distribute reprints for Government purposes notwithstanding any copyright notation herein.

{\it Competing Interests -- } 
Michael Hatridge serves as a consultant for D-Wave Inc. (formerly Quantum Circuits, Inc.), receiving remuneration in the form of consulting fees, and hold equity in the form of stock options.

\begin{acknowledgments}

\end{acknowledgments}
%\include{supplement}
%\nocite{*}
%\clearpage
%\bibliography{aipsamp}% Produces the bibliography via BibTeX.
\bibliographystyle{apsrev4-1}
\bibliography{refs.bib}
\end{document}

\renewcommand{\thefigure}{S\arabic{figure}} % Makes figure numbers S1, S2...
\appendix

\title{Supplement Material for: A superconducting quantum circuits single artificial atom maser}
\author{Maria Mucci}
\affiliation{Department of Physics and Astronomy, University of Pittsburgh, Pittsburgh, PA, 15260, USA}

\author{Nicholas Hougland}
\affiliation{Department of Physics and Astronomy, University of Pittsburgh, Pittsburgh, PA, 15260, USA}
\affiliation{Pittsburgh Quantum Institute, University of Pittsburgh, Pittsburgh, PA, 15260, USA}
\author{Chun-Che Wang}
\affiliation{Department of Applied Physics, Yale University, New Haven, CT, 06511 , USA}

\author{Israa Yusuf}
\affiliation{Department of Physics and Astronomy, University of Pittsburgh, Pittsburgh, PA, 15260, USA}

\author{Chenxu Liu}
\affiliation{Pacific Northwest National Laboratory, Richland, WA 99352, USA}

\author{David Pekker}
\email[email: ]{pekkerd@pitt.edu}
\affiliation{Department of Physics and Astronomy, University of Pittsburgh, Pittsburgh, PA, 15260, USA}
\affiliation{Pittsburgh Quantum Institute, University of Pittsburgh, Pittsburgh, PA, 15260, USA}
\author{Michael Hatridge}
\email[email: ]{hatridge@pitt.edu}
\affiliation{Department of Applied Physics, Yale University, New Haven, CT, 06511 , USA}
\maketitle

\section{Details of our theoretical model of the maser} 
\label{sec:Theory_Hamiltonian}
\begin{figure*}
    \centering
    \includegraphics[width = 0.9\linewidth]{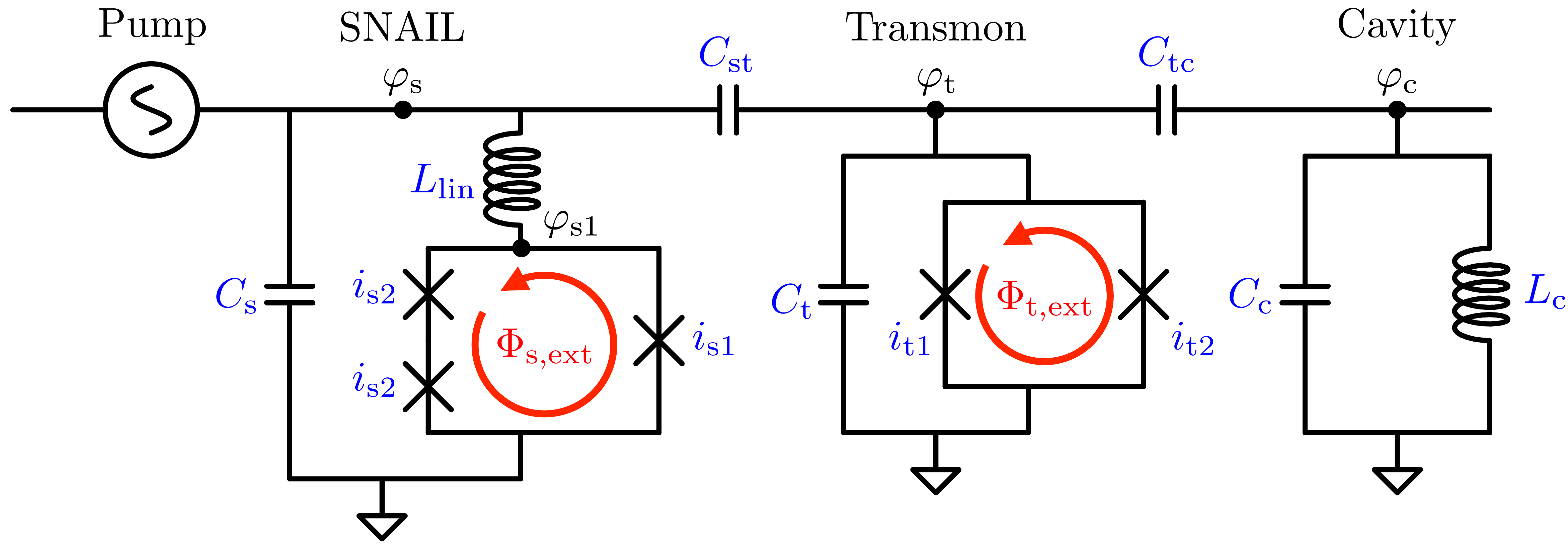}
    \caption{\label{fig:maserCircuit}Diagram of the maser circuit including the pump, SNAIL, transmon, cavity.}
\end{figure*}

We start modeling this system, using circuit QED methods~\cite{LesHouchesNotes_Girvin2014}, by writing the (classical) Hamiltonian of each of these three components (SNAIL, transmon, and cavity). Each component is shown in the circuit in Fig.~\ref{fig:maserCircuit}. We can write the Hamiltonian for the components as follows:
\begin{align}
    H_\snail &=\frac{1}{2} C_\snail \dot{\varphi}_{\snail1}^2-\phi_0 i_{\snail 1} \cos (\varphi_{\snail1} + \Phi_{\snail\text{, ext}}) \notag \\
    & - 2 \phi_0 i_{\snail 2} \cos (\varphi_{\snail1}/2) + \frac{1}{2 L_\text{lin}} (\varphi_\snail-\varphi_{\snail1})^2, \\
    H_\transmon &=\frac{1}{2} C_\transmon \dot{\varphi}_\transmon^2-\phi_0 i_{\transmon 1} \cos (\varphi_\transmon + \Phi_{\transmon\text{, ext}}) - \phi_0 i_{\transmon 2} \cos (\varphi_\transmon), \\
    H_\cavity &=\frac{1}{2} C_\transmon \dot{\varphi}_\cavity^2 + \frac{1}{2L_\cavity} \varphi_\cavity^2,
\end{align}
where $C_\alpha$ represent capacitances, $L_\beta$ represent inductances, $i_\gamma$ represent Josephson critical currents, $\Phi_{\delta,\text{ext}}$ represent external magnetic fluxes, and $\varphi_\epsilon$ represent dimensionless superconducting order parameter phase at the nodes of the circuit shown in Fig.~\ref{fig:maserCircuit}. We define $\varphi_\epsilon=\phi_\epsilon/\phi_0$, where $\phi_0=\frac{\hbar}{2e}$ is reduced magnetic flux quantum and $\phi_\epsilon$ is the flux. 

For the capacitive couplings between the components, we write the coupling Hamiltonians:
\begin{align}
    H_{\snail\transmon} &=\frac{1}{2} C_{\snail\transmon} (\dot{\varphi}_\snail-\dot{\varphi}_\transmon)^2\\
    H_{\transmon\cavity} &=\frac{1}{2} C_{\transmon\cavity} (\dot{\varphi}_\transmon-\dot{\varphi}_\cavity)^2.
\end{align}
Note that each of these coupling Hamiltonians includes terms which act on only a single component of the system, i.e. $\dot{\varphi}_\snail^2,\dot{\varphi}_\transmon^2,\dot{\varphi}_\cavity^2$. We treat these terms as part of the single-component Hamiltonians, leaving only the cross terms in the coupling Hamiltonians.
\begin{align}
    \tilde{H}_\snail &=H_\snail+\frac{1}{2} C_{\snail\transmon}\dot{\varphi}_\snail^2 \\
    \tilde{H}_\transmon &=H_\transmon+\frac{1}{2} (C_{\snail\transmon}+C_{\transmon\cavity})\dot{\varphi}_\transmon^2 \\
    \tilde{H}_\cavity &=H_\cavity+\frac{1}{2} C_{\transmon\cavity}\dot{\varphi}_\cavity^2 \\
    \tilde{H}_{\snail\transmon} &=C_{\snail \transmon}\dot{\varphi}_\snail \dot{\varphi}_\transmon\\
    \tilde{H}_{\transmon\cavity} &=C_{\transmon \cavity}\dot{\varphi}_\transmon \dot{\varphi}_\cavity.
\end{align}
Finally, we introduce quantum momentum operators by inverting the capacitance matrix 
\begin{align}
\begin{pmatrix}
C_{\snail}+C_{\snail \transmon} & C_{\snail\transmon} & 0 \\
C_{\snail\transmon} & C_{\transmon} + C_{\snail\transmon} + C_{\transmon\cavity} & C_{\transmon\cavity}\\
0 & C_{\transmon\cavity} & C_{\cavity} + C_{\transmon\cavity}\\
\end{pmatrix}
\end{align}
for the system. Using the inverse of the capacitance matrix, to first order in the couplings
\begin{align}
\begin{pmatrix}
\frac{1}{C_{\snail}}-\frac{C_{\snail \transmon}}{C_{\snail}^2} & -\frac{C_{\snail \transmon}}{C_{\snail} C_{\transmon}} & 0 \\
-\frac{C_{\snail \transmon}}{C_{\snail} C_{\transmon}} & \frac{1}{C_{\transmon}} - \frac{C_{\snail\transmon} + C_{\transmon\cavity}}{C_{\transmon}^2} & -\frac{C_{\transmon\cavity}}{C_{\transmon} C_{\cavity}}\\
0 & -\frac{C_{\transmon\cavity}}{C_{\transmon} C_{\cavity}} & \frac{1}{C_{\cavity}}-\frac{C_{\transmon \cavity}}{C_{\cavity}^2}
\end{pmatrix}
\end{align}
we write down the kinetic energy terms of the quantum Hamiltonian for the system
\begin{align}
    T_\snail &=-\frac{1}{2} \left(\frac{1}{C_{\snail}}-\frac{C_{\snail \transmon}}{C_{\snail}^2}\right) \partial_{\snail}^2 \\
    T_\transmon &=-\frac{1}{2} \left(\frac{1}{C_{\transmon}} - \frac{C_{\snail\transmon} + C_{\transmon\cavity}}{C_{\transmon}^2}\right) \partial_{\transmon}^2 \\
    T_\cavity &=-\frac{1}{2} \left(\frac{1}{C_{\cavity}}-\frac{C_{\transmon \cavity}}{C_{\cavity}^2}\right) \partial_{\cavity}^2 \\
    T_{\snail\transmon} &=\frac{C_{\snail \transmon}}{C_{\snail} C_{\transmon}} \partial_\snail \partial_\transmon\\
    T_{\transmon\cavity} &=\frac{C_{\transmon \cavity}}{C_{\transmon} C_{\cavity}} \partial_\transmon \partial_\cavity.
\end{align}
% $C_\snail \dot{\varphi}_\snail \rightarrow -i \partial_{\varphi_\snail}$, $C_\transmon \dot{\varphi}_\transmon \rightarrow -i \partial_{\varphi_\transmon}$, and $C_\cavity \dot{\varphi}_\cavity \rightarrow -i \partial_{\varphi_\cavity}$.

For each of the SNAIL, transmon, and cavity, we solve the Schrödinger equation to derive its eigenstates, and use these eigenstates as a basis to construct a Hamiltonian for the complete system. To do so, we discretize the phase variable and solve the resulting difference equations numerically (see Appendix \ref{sec:Derivs} for more details). Note that for the transmon and cavity, the phase Hamiltonian of each is dependent directly on the phase variable $\varphi_\transmon$ and $\varphi_\cavity$ directly. However, the SNAIL has the additional internal phase $\varphi_{\snail1}$ which determines how the phase $\varphi_{\snail}$ is divided between the SNAIL loop and a stray linear inductance $L_\text{lin}$. Handling of this internal phase and fitting the SNAIL to experimental data is discussed in Appendix~\ref{fig:snailFit}.
%we consider writing the wavefunction as a vector $\Psi$ which is discretized in phase. Then, it is possible to write the Hamiltonian of each component as a matrix, with derivatives computed using finite difference methods. See Appendix~\ref{sec:Derivs} for more details.

The energies of the eigenstates of each component can be used to produce a Hamiltonian using photon number in each component as a basis. The cross terms in the coupling Hamiltonians can also be constructed using these eigenstates by computing matrix elements for the $\tilde{H}_{\snail \transmon}$ and $\tilde{H}_{\transmon\cavity}$ terms, which can be done independently for each component since the operators for each component commute. That is, we can compute the matrix elements which go into the the coupling Hamiltonians in this basis as
\begin{align}
    \bra{\Psi_{j,n}} \partial_{\varphi,j} \ket{\Psi_{j,m}},
\end{align}
where $\Psi_{j,n}$ is the $n$-th eigenstate for the component $j=\snail,\transmon,\cavity$, and $\nabla_j$ is the finite differences matrix for the first derivative of that component (as defined in Appendix~ \ref{sec:Derivs}). We can then use this to write all necessary $\tilde{H}$ terms which will contribute to the system Hamiltonian for the three micromaser components and their couplings.

We write the contribution of the pump, which acts on the SNAIL directly, as
\begin{align}
    H_{\text{pump}} = i \Omega \cos(\omega_\text{p} t) \dot{\varphi}_\snail,
\end{align}
where $\Omega$ is the amplitude of the pump and $\omega_\text{p}$ is the frequency of the pump.

As the SNAIL-transmon coupling is much stronger than the transmon-cavity coupling, we proceed by diagonalizing the artificial atom Hamiltonian $H_{\atom}=\tilde{H}_\snail+\tilde{H}_{\transmon}+\tilde{H}_{\snail\transmon}$. In the eigenbasis of the artificial atom Hamiltonian, the basis states are nearly product states of the original basis with a fixed photon number on each component -- which allows us to label the new eigenstates by photon numbers.  It is in this new basis that we apply Lindbladian loss which affects the SNAIL and transmon. Specifically, SNAIL photon loss is applied between eigenstates of the artificial atom which differ by one photon in the SNAIL (according to their most similar states in the original basis). The same is true for transmon photon loss, which is applied between eigenstates which differ by one photon in the transmon (according to their most similar states in the original basis). This scheme for applying the Lindbladian terms reduces the effect of Purcell loss,  the origin of which is discussed in Ref.~\cite{PurcellEffect_Purcell1946, MultimodePurcell_Houck2008, WISPE_Patel2025}. We apply these losses in an uncorrelated way. I.e., we consider a Lindbladian associated with the loss between any two states independently, rather than a single Lindbladian operator encompassing all SNAIL or transmon loss processes. The reason for treating the different loss processes as incoherent is that the photon frequencies associated with the different loss processes are different from each other because of the nonlinearities of the artificial atom subsystem. On the other hand, for the cavity loss, we consider a single Lindbladian with correlated losses, which are between states differing by one cavity photon. 
For the pump, we take $H_\text{pump}$ and transform it into the eigenbasis of the artificial atom. We keep only the terms that are relevant, that is the terms which differ by 1 photon in the SNAIL and 1 photon in the transmon~\cite{ChemPot_Mucci2025}. This includes, for example, pumping from the ground state of the artificial atom $\ket{0, g}$ to the $\ket{1, e}$ state, or from the $\ket{1, e}$ to $\ket{2, f}$ state, and so on. It also includes pumping from $\ket{0, e}$ to $\ket{1, f}$, $\ket{1, g}$ to $\ket{2, e}$, and similar contributions. For each state that we keep, we will find the maximum magnitude of the matrix element (when $t=0$) and incorporate the rotation as $\exp(i \omega_{\text{p}} t)$. We refer to this new, reduced pump as $\tilde{H}_\text{pump}$.
%That is, after $H_\text{pump}$ is transformed into the new basis, the matrix elements of the new pump are
% \begin{align}
%     \tilde{H}_{\text{pump},nm}=|H_{\text{pump},nm}(t=0)|\exp(i \omega_{\text{pump}} t).
% \end{align}
We then write the master equation of the system,
\begin{align}
\begin{aligned}
    \dot{\rho}=&-i[H_\atom+\tilde{H}_\cavity+\tilde{H}_{\transmon \cavity}+\tilde{H}_{\text{pump}},\rho] \\
    +&\chi_\snail\sum_{i=1}^{n_\snail} \big(\hat{\snail}_i\rho \hat{\snail}_i^\dagger-\frac{1}{2} \hat{\snail}_i^\dagger \hat{\snail}_i \rho - \frac{1}{2} \rho \hat{\snail}_i^\dagger\hat{\snail}_i\big)\\
    +&\chi_\transmon\sum_{i=1}^{n_\transmon} \big(\hat{\transmon}_i\rho \hat{\transmon}_i^\dagger-\frac{1}{2} \hat{\transmon}_i^\dagger \hat{\transmon}_i \rho - \frac{1}{2} \rho \hat{\transmon}_i^\dagger\hat{\transmon}_i\big)\\
    +&\chi_\cavity \big(\hat{\cavity}\rho \hat{\cavity}^\dagger-\frac{1}{2} \hat{\cavity}^\dagger \hat{\cavity} \rho - \frac{1}{2} \rho \hat{\cavity}^\dagger\hat{\cavity}\big),
\end{aligned}
\label{eq:MasterEquation}
\end{align}
where $\rho$ is the density matrix for the system and $\chi_j$ is the decay rate for the component $j$ (SNAIL, transmon, cavity). The operators $\hat{s}_i$ and $\hat{t}_i$ and their Hermitian conjugates are the creation and annihilation operators which induce transitions between eigenstates in the artificial atom basis where the SNAIL and transmon photon numbers change by one. We have separate operators for each pair of states of the artificial atom in order to implement uncorrelated losses. The number of states in our Hilbert space for each of these components is $n_\snail$ for the SNAIL and $n_\transmon$ for the transmon. For the cavity, we write the standard creation and annihilation operators as $\hat{c}^\dagger,\hat{c}$ since losses are correlated ($n_\cavity$, the number of states we consider in the cavity, does not appear in the expression above for this reason). We have ensured that our choices of $n_\snail,n_\transmon,n_\cavity$ are sufficient for the conditions we simulate by checking that our results do not vary significantly as we increase these cutoffs.

This master equation can be considered as a superoperator $\mathcal{L}$ for $\dot{\rho}$ acting on $\rho$, i.e. $\dot{\rho}=\mathcal{L} \rho$. However, this superoperator contains rotating terms due to the pump, which makes analysis of the system more complex. To simplify the problem, we transform to a rotating basis and drop the remaining rotating terms, as described in Appendix \ref{sec:rotate}. Then, the steady state of the maser can be determined by computing the eigenstate of $\mathcal{L}$ which has eigenvalue 0. The next smallest eigenvalue by real part gives the linewidth of the system. Therefore, constructing and analyzing $\mathcal{L}$ allows us to determine maser behavior which we can sweep over experimental parameters such as transmon and pump frequencies.

\section{Details of implementing the circuit QED maser model} 
\label{ModelDetails}

\subsection{Derivative operators} 
\label{sec:Derivs}

\begin{figure}
    \centering
    \includegraphics[width = 0.9\linewidth]{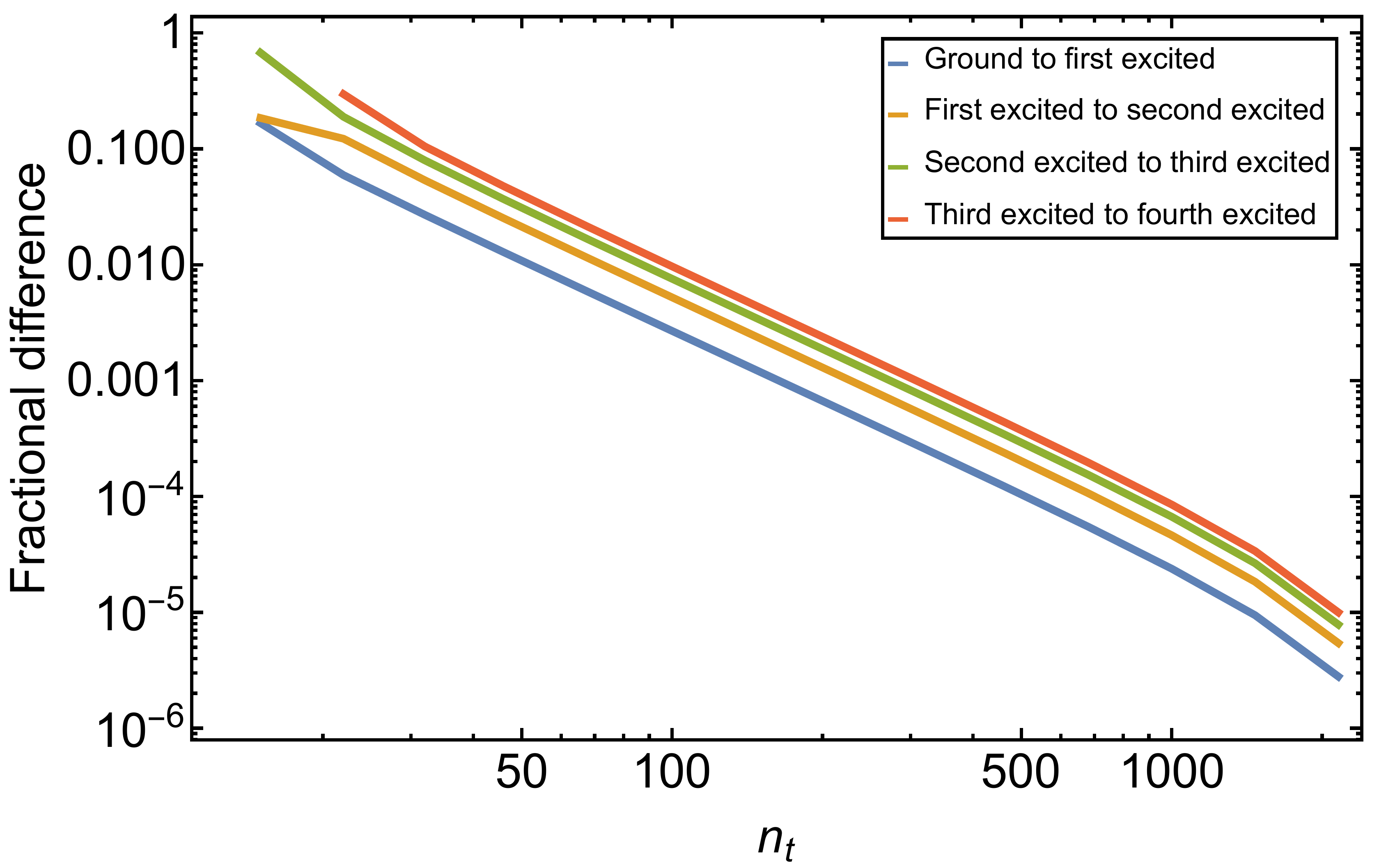}
    \caption{Convergence of spacing of the levels of the transmon. For various $n_\transmon$, we plot the spacings between the 5 lowest levels to show how they converge as $n_\transmon$ increases. Here, we plot the fractional difference between these spacings and the spacings computed with $n_\transmon=3000$. By $n_\transmon=1000$, the difference is smaller than one part per 10000, and so we consider this sufficient. }
    \label{fig:levelConvergence}
\end{figure}

Upon quantization, the momentum operator $C \dot{\varphi}$ becomes $-i \hbar \nabla_\varphi$ (and we use the units where $\hbar=1$). In order to implement derivatives in the discrete Hamiltonian, we use the finite differences method to define operators for first and second derivatives for each component. Given the value of a function $\Psi$ at a point $\varphi_0$ and its nearest neighbors in a grid with spacing $h_\varphi$, we express the first and second derivatives at the point as
\begin{align}
f'(\varphi_0)=&\frac{-f(\varphi_0-h_\varphi)+f(\varphi_0+h_\varphi)}{2h_\varphi} \\
f''(\varphi_0)=&\frac{f(\varphi_0-h_\varphi)-2f(\varphi_0)+f(\varphi_0+h_\varphi)}{h_\varphi^2}.
\end{align}
Therefore, the discrete first derivative operator with respect to $\varphi$ for each component (SNAIL, transmon, cavity) can be expressed as
\begin{align}
\nabla_{\varphi}=
\frac{1}{2 h_{\varphi}}
\begin{pmatrix}
0 & 1 & 0 & 0 & \dots\\
-1 & 0 & 1 & 0 & \dots\\
0 & -1 & 0 & 1 & \dots\\
0 & 0 & -1 & 0 & \dots\\
\vdots & \vdots & \vdots & \vdots & \ddots\\
\end{pmatrix}
\end{align}
where for each component we select an appropriate spacing $h_\varphi$. For the cavity, we use $n_\cavity=14000$ and $h_{\varphi,\cavity}=0.0003$. This small spacing is necessary to preserve the harmonic nature of the cavity. We have confirmed that this $n_\cavity$ is sufficient by verifying that the spacings of the cavity states are the same to within $2$~kHz, which is small compared to the spacings themselves which are on the order of $7$~GHz. For the transmon and the SNAIL, the total range of possible phases is fixed (to $2\pi$ for the transmon and $4\pi$ for the SNAIL due to the SNAIL design). We select $n_\transmon=1000$ and $n_\snail=1000$, fixing the spacing of points for each. Since the transmon and SNAIL are periodic, we set the top right and bottom left elements in the first derivative matrix to -1 and 1, respectively, since the points at 0 and $2\pi$ are equivalent. We build the second derivative matrix for each component in a similar manner. We confirm that $n_\snail$ and $n_\transmon$ are sufficient since the energies of their first five levels do not change significantly by increasing these parameters further, as show in Fig.~\ref{fig:levelConvergence}.

\begin{figure}
    \centering
    \includegraphics[width = 0.9\linewidth]{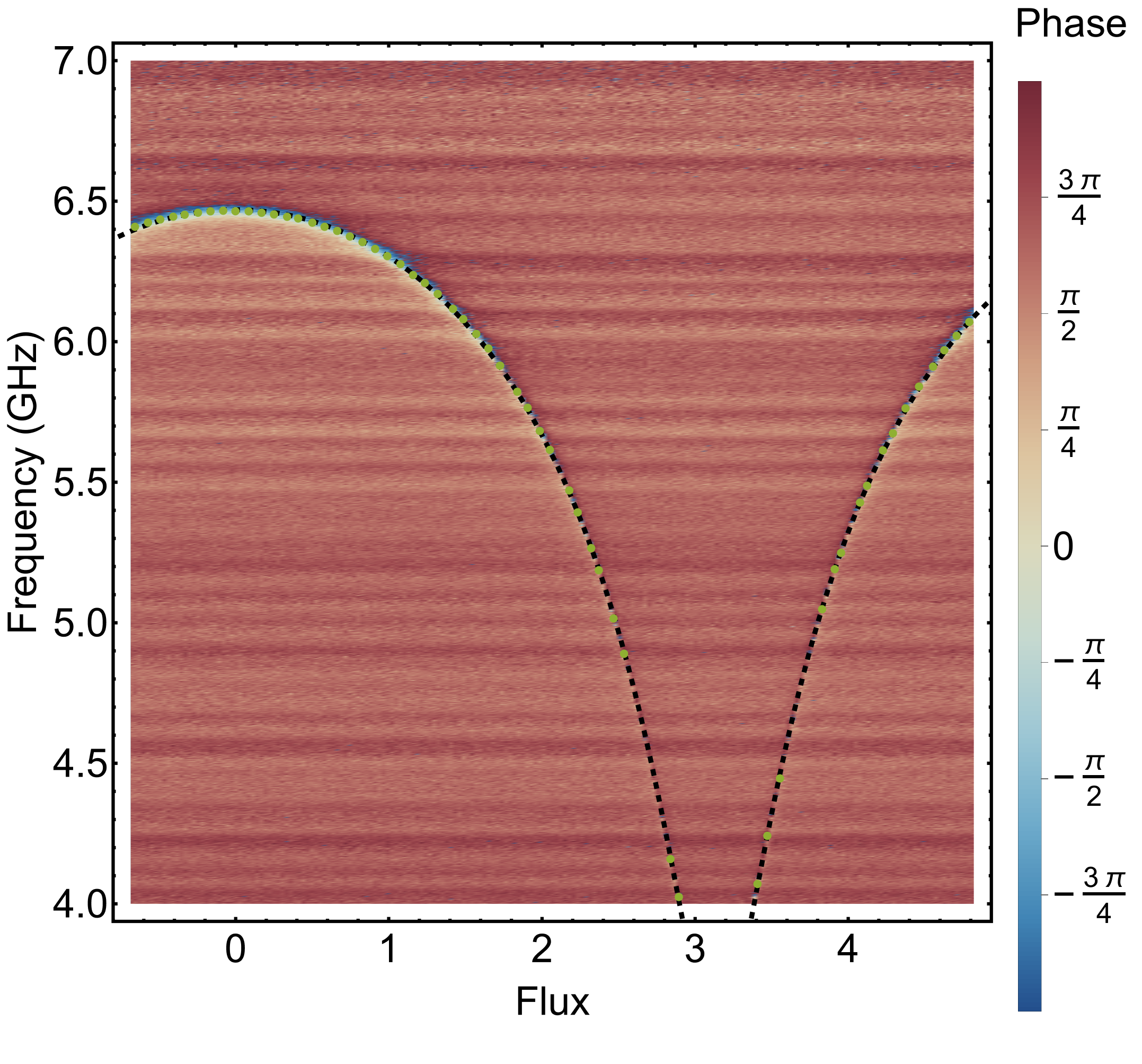}

    \caption{Plot of the ground to first excited state transition of the SNAIL. The green points are extracted from the experimental data and lie along the transition over the flux sweep. The black dotted line shows the frequency of the transition according to our theory after fitting.}
    \label{fig:snailFit}
\end{figure}

\subsection{Fitting and modeling the SNAIL} 
\label{snailFit}

Modeling our SNAIL, like modeling the transmon and cavity, requires fitting its circuit parameters such that the SNAIL behavior matches experimental data. However, the SNAIL is more complicated than the other components because we include in our model a stray linear inductance $L_\text{lin}$ which divides the phase $\varphi_\snail$. Therefore, the SNAIL is dependent not only on $\varphi_\snail$, but also on $\varphi_{\snail1}$ which is itself dependent on $\varphi_\snail$, as it is the phase which minimizes the energy of the SNAIL component for a given value of $\varphi_\snail$. In discretizing the phase variable for the SNAIL, the phase $\varphi_{\snail1}$ must be determined for each desired value of $\varphi_\snail$.

With this intermediate phase considered in our model, we have 3 fitting parameters to determine, $i_{\snail1},i_{\snail2}, \text{ and } L_{\text{lin}}$. The capacitance $C_\snail$ is inferred by experimental data to be $341$~pF.  To find the values of the fitting parameters, we fit the ground state to first excited state transition to experimental data as shown in Fig.~\ref{fig:snailFit}.

\subsection{Applying a rotating frame and dropping rotating terms} \label{sec:rotate}

The master equation as written in Eq.~\ref{eq:MasterEquation} contains rotating terms due to the pump. We apply a rotation which takes us to a frame where these rotating terms are stationary which allows for tractable analysis of the master equation without integrating in time to find the steady state solution. In particular, we apply the unitary transformation which rotates the system based on the number of photons in the SNAIL. That is, each state is rotated by $e^{i \snail \omega_\text{p} t}$, where $s$ is the number of photons in the SNAIL (in the transformed basis where SNAIL and transmon are hybridized). We apply the rotation in this way since the pump acts on the SNAIL, inducing a rotation at $\omega_\text{p}$. This counter-rotation applies to terms in the pump which link two states with different SNAIL photon number, with the counter-rotation always having a net effect of rotating by $-\omega_\text{p}$. The result is that the pump, when written in the rotating frame, contains only stationary terms.  
However, any rotation will induce new rotating terms in $\tilde{H}_{\transmon\cavity}$, the transmon cavity coupling. The rotation we have chosen is advantageous since the most important terms in this coupling do not couple states with different SNAIL number, and thus will not be rotating when applying our transformation. After applying our rotation, we drop all remaining rotating terms in $\tilde{H}_{\transmon\cavity}$ and arrive at a new Hamiltonian with no rotating terms which closely approximates the original Hamiltonian.

\section{Experiment Details}
\label{Experiment_setup}
\subsection{Device setup and mode details}

\begin{figure*}
\centering
\includegraphics[width = 1\linewidth]{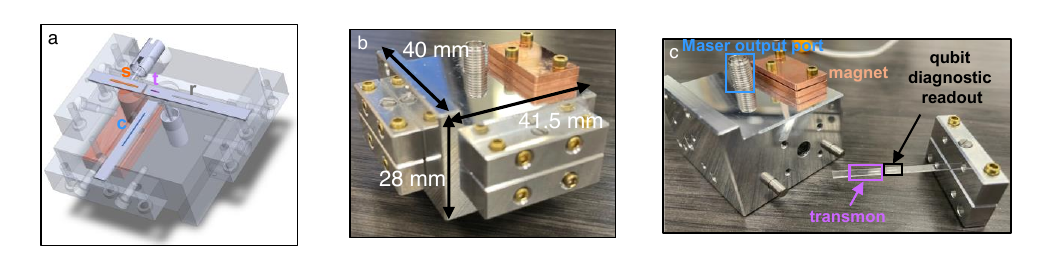}
\caption{\label{fig: device's photos} Photo of the maser device. (a) Cross section of the 3D Auto CAD drawing of the device (b) The fully assembled housing with its dimensions. (c) The chip holder clips are aligned to the housing with Titanium dowel pins. All the sapphire chips are inserted into the extruded tubes.}
\end{figure*} 

The device is made of superconducting thin-film elements on sapphire substrates in a superconducting aluminum housing pictured in Fig.~\ref{fig: device's photos}. This housing is a block of 6061 aluminum with extruded small tubes that are spanned by sapphire chips~\cite{Tube_Axline2016}. The outer surfaces of the aluminum were mirror polished to ensure clean mating seams and surfaces between the housing and its connection to the mixing chamber plate. There are also extrusions in the housing block for SMA microwave coupling pins to be held at the needed distance from each circuit element, which allows for routing microwave signals in and out from room temperature for control and readout. 

The large features of both the SNAIL and transmon are made of sputtered tantalum, and the small Josephson junctions made of aluminum/aluminum oxide/aluminum in the bridge-free 'Manhattan' style~\cite{ManhattanJJs_Potts2001}. The SNAIL has two large junctions in parallel with one small one at the inductive energy ratio $\alpha_s = 0.3$~\cite{SNAIL_Frattini2017}. The transmon has two junctions of equal size in parallel. Both components of our artificial atom require separate flux biasing to thread flux through their respective Josephson junction loop. This was achieved with creating two large openings in the aluminum housing above each element, fitting a copper cup into the opening, and placing a solenoid magnet inside the cup. The magnets are made from CuNi clad superconducting NbTi wires spun around a copper spindle, with leads on either end connecting to DC wiring in the fridge that goes to a room temperature current source (YOKOGAWA GS200). One of these copper cup openings in the aluminum go halfway through the housing, and the other completely through. This ensures a unique path for flux to travel and creates independent magnet control (a flux bias applied to either the SNAIL or transmon should not effect the other's bias). Without the asymmetric tube openings, flux biasing was completely coupled between the two. With it, we achieved minimal cross-talk and only required small current corrections to maintain desired frequencies at combinations of SNAIL and transmon biases.

The SNAIL was designed to be low Q so it could be reset quickly and act as the dump mode in our photon ratchet. Its operable frequency ranged from $4.5-6.5$~GHz, but we want to operate it at a point of high $g_{sss}$ and low $g_{ssss}$, or high odd-order nonlinearity and low even-order~\cite{KerrFreeSNAIL_Sivak2019}. Near our best masing point (Fig.~\ref{fig:Lowest_BW}), $f_s = 5.76$~GHz and $\kappa_s/2\pi = 24.5$~MHz. The operating $g_{sss}$ is designed to be $\approx -100$~MHz. We may tune the SNAIL to a point of higher even-order nonlinearity, like at $f_s= 6.0$~GHz which sits closer to the maximum of Fig.~\ref{fig:snailFit}.

The transmon is a DC SQUID whose frequency follows $f_t \propto \sqrt{|\cos{\frac{2 \pi \Phi}{\Phi_0}|}}$. We may only measure in the range of $\approx 6.9-7.2$~GHz due to the current bias. During maser operation, the transmon and cavity are brought into resonance. The cavity is fixed at $6.971$~GHz when far-detuned from the transmon. 

The direct coupling $g_{tc}/2\pi = 0.44$~MHz is large compared to the transmon-cavity detuning (which should be $\Delta_{tc} = 0$ during operation), and so we cannot dispersively measure the qubit through the masing cavity~\cite{DispersiveReadout_Wallraff2005}. To probe the qubit, purely for diagnostic purposes, we strongly couple the transmon to a low Q cavity mode, $r$, for readout. This mode is far detuned from the maser mode at $f_r = 8.547$~GHz and is not involved in the parametric gain pump. The $r$ mode plus a weakly coupled qubit microwave drive pin allow us to perform QND dispersive measurements for values like qubit lifetimes. The microwave coupling pin associated with this auxiliary $r$ mode was placed using the frequency-based argument of WISPE~\cite{WISPE_Patel2025} where the resonator pin is placed at a point of zero qubit field.

\subsection{Cryogenic and room temperature measurement setup}

\begin{figure*}
\begin{center}
\includegraphics[scale = 0.5]{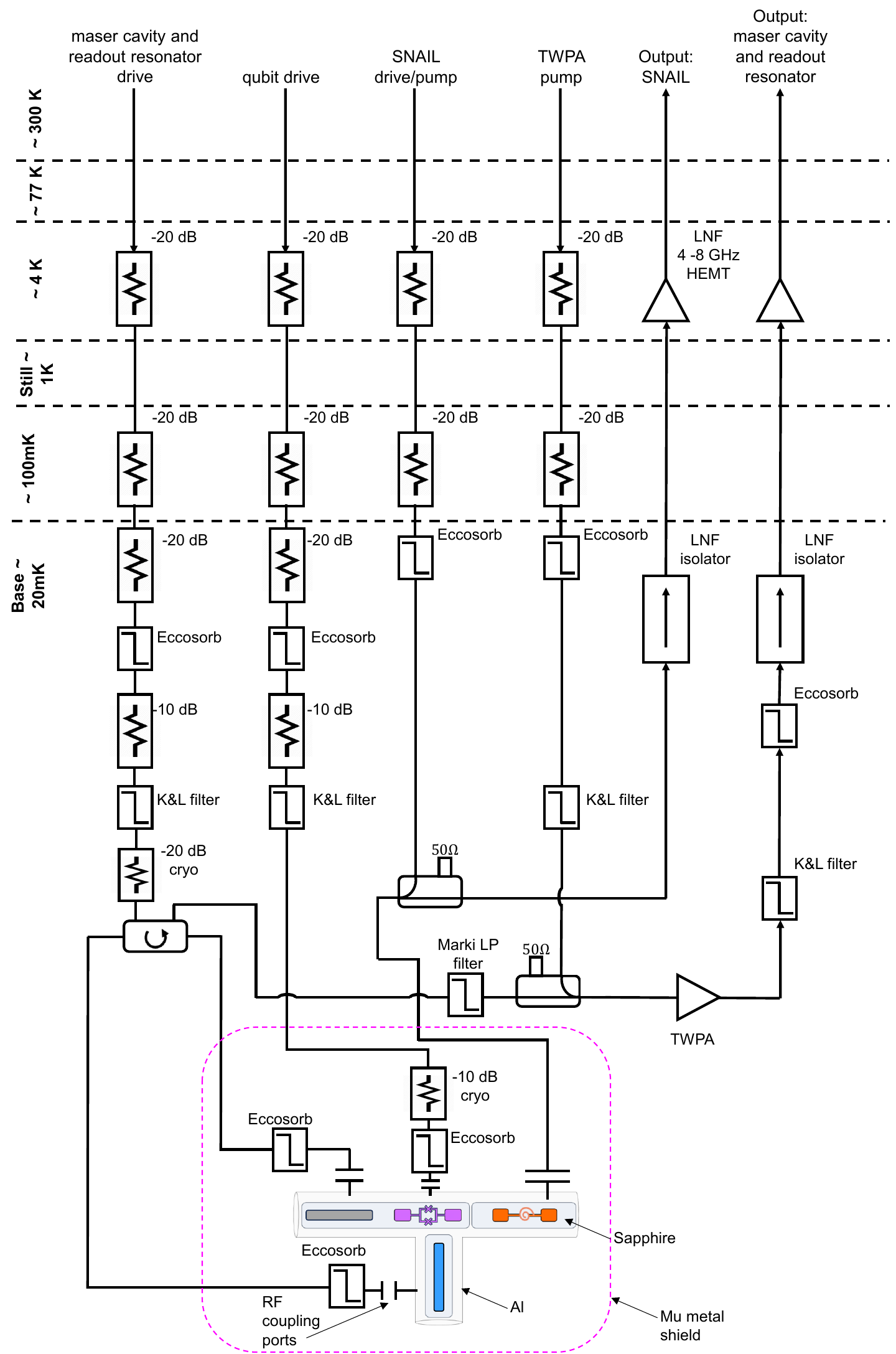}
\caption[\textbf{Cryogenic diagram}]{\textbf{Cryogenic diagram} A diagram of the full cryogenic input and output setup for our maser. The masing cavity and readout resonator are chained on the same input\slash output chain because they are spectrally separated and never simultaneously measured. Transmon drives and SNAIL readout are only performed for diagnostics. During the masing process, only two pumps are driven: the TWPA and parametric gain down the SNAIL line. The masing cavity is read on its output line.}
\label{fig:cryo_diagram}
\end{center}
\end{figure*}

Measurements of the SNAIL $s$, the weakly probed bare cavity $c$, and transmon $t$ through the readout mode $r$ were taken in reflection. The output of the maser was measured as signal output to the data collection device (either the spectrum analyzer or QICK ADC). The cryogenic setup of our device, mounted at the base stage of a cryogen-free dilution refrigerator, is shown in Fig.~\ref{fig:cryo_diagram}. During operation of the maser, there are only two input drives: the parametric pumps on the SNAIL and the TWPA drive. The only data collected is through the masing cavity output port. A TWPA provided to us through MIT Lincoln Laboratory~\cite{TWPApaper_Macklin2015} backs the masing cavity to increase our SNR and allow for faster measurement, as increasing measurement time far beyond the coherence time of the maser would obfuscate its coherent properties. 

\begin{figure*}
\begin{center}
\includegraphics[scale = 1.0]{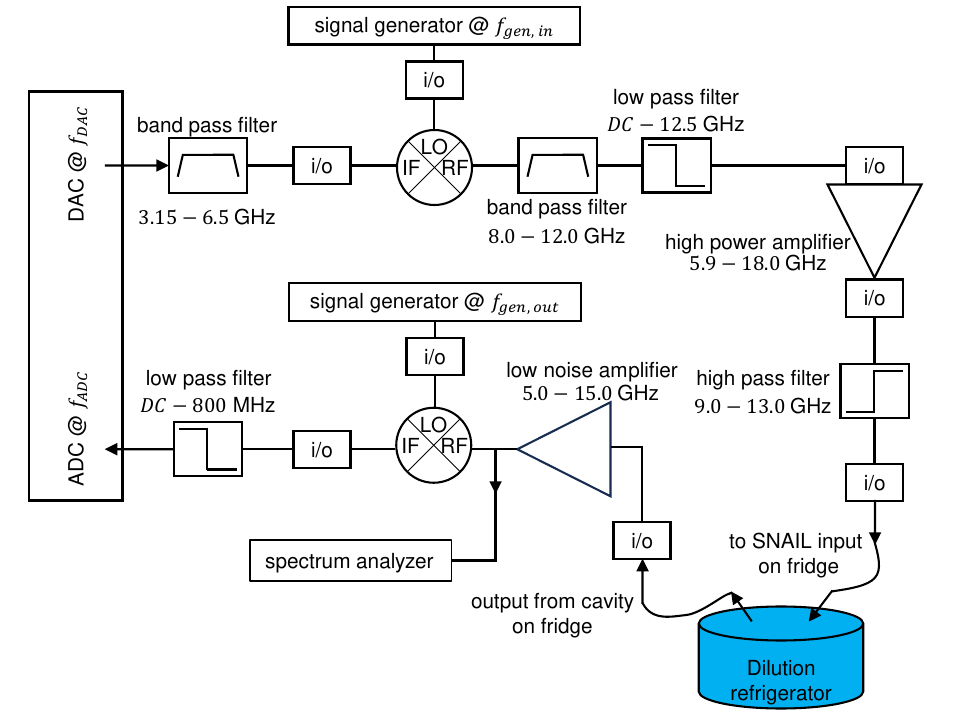}
\caption[\textbf{Room temperature microwave electronics}]{\textbf{Room temperature microwave electronics} The input for the parametric drive down the SNAIL input line, and output from the masing cavity on its dedicated output line. Signals are generated by the DAC at $f_{DAC}$, which mixes with an LO $f_{gen, in}$ to send frequency into the fridge at $ f_{gen, in} - f_{DAC} = f_d = f_s + f_t + \gamma_d$. There are a series of filters to prevent unwanted sidebands and AWG aliasing tones from going into the fridge, which can potentially interfere with the intended processes. To read out the signal from the cavity, we can choose two options. We can measure broad band responses on the spectrum analyzer, and\slash or can we digitize the signal by mixing the cavity output signal $f_{c}$ with generator tone $f_{gen, out}$ to reach the ADC at $f_{c} - f_{gen, out} = 90$~MHz.}
\label{fig:M_RT_interferometer}
\end{center}
\end{figure*}

At room temperature, we use two methods to generate and read signals from our quantum system. Weakly probed signals are connected to a VNA (Keysight P9374A), with a direct connection of one side of the VNA into the $300$~K input ports indicated at the top left of Fig.~\ref{fig:cryo_diagram}, and the other side of the VNA to the corresponding output ports on the right of Fig.~\ref{fig:cryo_diagram} through a room temperature low noise amplifier (Mini-Circuit ZX60-14LN-S+). Two tone spectroscopy was performed on the transmon using an additional signal generator (SignalCore SC$5511$A) routed to the transmon's input port. An additional signal generator (Keysight Agilent N5183B) was used to supply a constant excitation to the TWPA input port. For frequency domain maser measurements (described first in Section~``\nameref{sec:data_section}''), the maser cavity's output port was connected to a spectrum analyzer (Keysight N9020A MXA) through the room temperature low noise amplifier. The parametric pump was supplied through the QICK RFSoC 111 DAC, mixed with the LO of a SignalCore generator, and through a series of filters and amplifiers sent to the SNAIL's input port. For the time resolved measurements (described second in Section~``\nameref{sec:data_section}'', the output from the maser cavity provided the RF tone, a SignalCore the LO, and the mixed signal out of the IF port of a mixer (Marki IRW0618) was sent to a QICK ADC channel for readout down-converted to $90$~MHz. The room temperature microwave electronics set up for the masing procedure is shown in Fig.~\ref{fig:M_RT_interferometer}.

We found that the pump and bias conditions used to induce masing were not consistent with time, meaning we would return to some particular combination of conditions (within the same fridge cooldown cycle) and find the maser's output behavior different than it had been at some previous time. This can be partially attributed to transmon frequency shifting vs. time. If the qubit changes frequency, it effectively induces an unknown detuning on the pump frequency because we cannot measure the transmon directly to detect this frequency shift while it is used in the masing process. It can also be attributed to unstable control and readout electronics. If the output of our electronics drifts with time and external conditions (like the temperature of the room), then our measurements can only be as coherent as this instability.

All of the room temperature components shown in Fig.~\ref{fig:M_RT_interferometer} must have commensurate frequencies and timing, and remain stable in their output or signal passing with time. All components, when possible, are locked to the same $10$~MHz Rubidium frequency standard so the generators and spectrum analyzer `agree' on what a Hertz is. Additionally, we took great care to stabilize the temperature of every component so their output is consistent. The SignalCore generators, amplifiers, mixers, and filters were bolted to a rack mount box with optical board inside, and the box was insulated with Styrofoam \cite{Subharmonics_Xia2025}. Inside the box was a temperature sensor read through a Raspberry Pi, and outside the box was a water-cooled Peltier cooler. We set the temperature that we wanted the box inside, and the Pi used a PID loop to control the cooler, which is temperature controlled with chilled water. We tracked the output of the drive line vs. time and find it stable to within $1\%$ over roughly an hour. This method better guaranteed that the bandwidth of the signal emitted by the masing cavity, on the tens of Hertz level, was not limited by room temperature electronics. 

\section{Measurement Details}
\label{Msmt_details}

\subsection{Data fitting}
\label{data_fitting}
The masing cavity emits light that can be captured by either a spectrum analyzer, capturing slow steady-state emission, or with a digitizing ADC which captures fast, time-resolved snapshots of emission. On the spectrum analyzer, the light peak is a coherent state with a Lorentzian shape. We extract the frequency of the cavity when masing, $f_c^{masing}$, its power output $P_c^{masing}$, and its full width at half maximum (FWHM) to fit for the linewidth of the peak using the formula:
\begin{align}
    \label{Eq: Lorentzian}
    \mathcal{L} = \frac{1}{\pi} \frac{\Gamma}{2} \frac{1}{(f - f_0)^2 + \frac{\Gamma}{2}^2)} 
\end{align}
\noindent for a Lorentzian with bandwidth $\Gamma$ and center frequency $f_0$. This gets scaled by:
\begin{align}
    \label{Eq: Lorentzian_fit_scaled}
    \mathcal{L}_{\text{linear}} = A \mathcal{L} + c 
\end{align}
\noindent to fit the Lorentzian power spectrum in linear units, as our analyzer produces data in dBm. This fit was used throughout the experiment and produces the fit curve and extracted bandwidth $\Gamma_c^{masing}$ in Fig.~\ref{fig:Lowest_BW}b.

For digitized readout, we use the two-time correlation function to track the maser's output. The radius of the IQ plane ring produced in Fig.~\ref{fig:TimeRings_narrowest}
% ~\ref{fig:TimeRings_narrowest}
remains constant, barring its finite thickness due to noise, meaning we may use the coherence of successive measurements to determine its decay lifetime. We use the function:
\begin{align}
    G(t) = \langle c^{\dagger}(t') c(t) \rangle \sim e^{\omega_c \abs{t' - t}}
\end{align} 
to find how IQ pairs at time $t'$ compare to those at time $t$. When unwrapped around $\pm 2\pi$ this follows an exponential decay curve, fit for coherence time or linewidth. This method produces values that agree with our slow measurements, but offers greater insight into the behavior of the maser as it evolves. While the spectrum analyzer method requires less tuneup and precise knowledge of the maser's frequency, it does not show us ring-down time or how the finite-width Lorentzian peak is formed.

\subsection{Maser Tuning}
\label{tuneup}

\begin{figure*}
\begin{center}
\includegraphics[scale = 1.0]{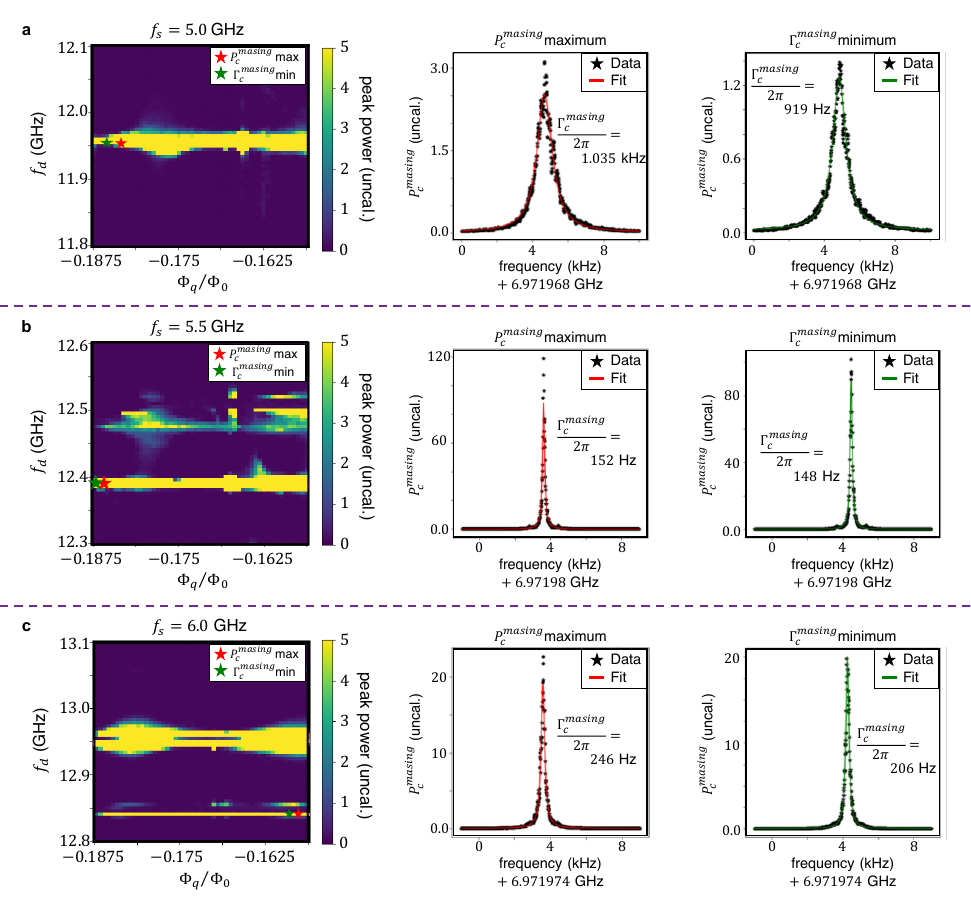}
\caption{\textbf{Masing vs. SNAIL bias, transmon bias, pump frequency} In each colorplot, we vary flux through the transmon magnet ($\Phi_q$) at a fixed $f_s,~g_{sss},~g_{ssss}$, and pump power. We also vary the pump frequency $f_p$. We measure the cavity's output on the spectrum analyzer, and note bright emissions across biases. The maximum brightness output in the sweep $P_c^{masing}$ is marked in red and the minimum bandwidth in the sweep $\Gamma_c^{masing}$ marked in green. Top to bottom we vary SNAIL bias to change its frequency and nonlinear coefficients (a) $f_s = 5.0$~GHz (b) $f_s = 5.5$~GHz, and (c) $f_s = 6.0$~GHz.}
\label{fig:M_masing_vs_snail}
\end{center}
\end{figure*} 

One of the key features of this device is its wide operation window. The transmon may be detuned from the masing cavity and the SNAIL may be both detuned in frequency and in nonlinearity from its ideal $g_{ssss} = 0$ operating point. Additionally, the driving amplitude of the SNAIL's parametric pump adds another control knob to modulate the maser's performance. 

There are a few main reasons to sweep parameters. First, no one control knob produces effects on only one parameter. Second, we expected detunings in the system due to factors like hybridization, fabrication imperfections, strong drives, and drifts vs time. Additionally, the broadband parametric mixing element key to this experiment introduces variability and flexibility. 

\begin{figure*}[!h]
\begin{center}
\includegraphics[scale = 0.6]{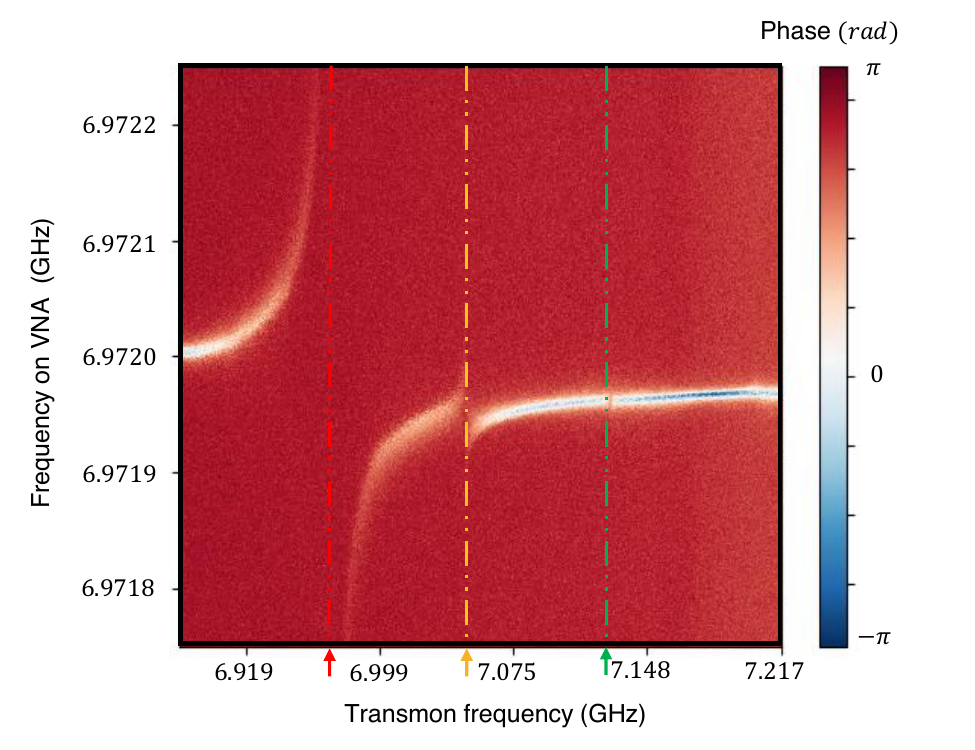}
\caption{\textbf{Maser cavity response vs. transmon frequency bias} Sweeping transmon frequency while measuring maser cavity on VNA. The avoided crossings indicate that the transmon transitions $\ket{e} \rightarrow \ket{g}$ (red), $\ket{f} \rightarrow \ket{g}$ (here half the frequency, yellow), and $\ket{f} \rightarrow \ket{e}$ (green) are on resonance with the maser cavity. The avoided crossing match with the maser bright spots shown in Fig.~\ref{fig:maserLevels}.}  
\label{fig:Maser spectroscopy}
\end{center}
\end{figure*} 

\begin{figure*}[!h]
\begin{center}
\includegraphics[scale = 1.0]{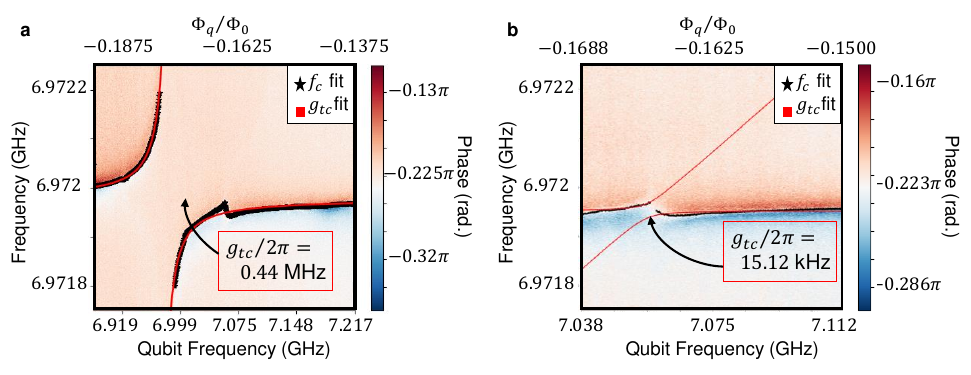}
\caption[\textbf{Avoided crossing: qubit and cavity}]{\textbf{Avoided crossing: qubit and cavity} Weakly measuring the cavity while sweeping the transmon's flux brings the qubit into resonance with the cavity. When they hybridize, their coupling Hamiltonian (see Sec.~\ref{sec:Theory_Hamiltonian}) is rediagonalized with one eigenvalue shifted by up $g_{tc}$ and the other down by $g_{tc}$. (a) When the qubit's $f_{ge, \text{bare}} = f_{c, \text{bare}}$ they couple with strength $\frac{g_{tc}}{2\pi} = 0.44$~MHz. (b) When $f_{gf / 2, \text{bare}} = f_{c, \text{bare}}$ they couple with strength $\frac{g_{tc}}{2\pi} = 15.12$~kHz.}
\label{fig: Avoided crossing fits}
\end{center}
\end{figure*} 

The first parameter swept in our experiment is the transmon bias current which tunes its frequency by threading external flux through its parallel loop of Josephson junctions. This allows us to bring the transmon into resonance with the maser cavity and facilitate the direct exchange interaction between them that ratchets photons from the qubit to the cavity. Figure~\ref{fig:Maser spectroscopy} shows the phase component of single-tone spectroscopy taken of the maser cavity on a VNA. The current through the transmon bias magnet is swept, corresponding to varying the transmon frequency along the x-axis. For every input tone frequency (y-axis), the phase response of the mode is measured (colorbar). We find an avoided crossing for the primary transition $\ket{e} \rightarrow \ket{g}$ with a direct coupling $g_{tc}$ fit to $0.44$~MHz (Fig.~\ref{fig: Avoided crossing fits}(a)). The coupling rate $g_{tc}$ is measured as one half of the vertical split at the resonance divergence point. Across many cooldowns we found secondary and tertiary transmon-maser crossings that correspond to bias conditions where the negatively-detuned higher order transitions of the transmon are what cross the cavity's frequency. In Fig.~\ref{fig: Avoided crossing fits}b we fit the avoided crossing for the $\ket{f}  \rightarrow \ket{g}$ transition (marked in yellow in Fig.~\ref{fig:Maser spectroscopy}) to $15.12$~kHz. 

It is important to point out that the rate hierarchy of Eq.~\ref{eq:hierarchy} is satisfied by the $f_{ge}$ crossing at rate $g_{tc}/2\pi = 0.44$~MHz with $\Gamma_c/2\pi = 6-20$~kHz. There is variance in this value due to qubit-cavity hybridization causing varying cavity lifetimes. However the hierarchy is not always satisfied with the $f_{fe}/2$ crossing $=15.12$~kHz. This means that at this bias, the cavity will become decoherent at a similar or faster rate than light can be supplied to it by the inverted qubit. Our bath engineering protocols mean we `fight' the natural rates of the system to sustain masing \cite{ChemPot_Mucci2025}.

Figure~\ref{fig:M_masing_vs_snail} shows one of the parameter sweeps performed on the maser: In each panel top to bottom, a-c, we park the SNAIL at a fixed frequency and sweep both the transmon frequency (x-axis on the color plot) and pump frequency (y-axis on the color plot) at a set pump power. The individual spectrum analyzer peaks shown to the right of each color plot correspond to the maximum power from that 2D sweep (red fit) and minimum bandwidth (green fit). The SNAIL frequency does change slightly with changing the transmon's bias, necessitating a flux correction applied to the SNAIL for particular combinations of transmon applied flux and desired SNAIL frequency. The SNAIL's properties, including its nonlinearity, have a large impact on maser performance because it scales the parametric up rate $\Gamma_{ge}^{p}$ which must overcome transmon decay and be faster than the exchange rate of photons from the transmon into the cavity. The three chosen SNAIL biases in Fig.~\ref{fig:M_masing_vs_snail} do overcome this rate scaling, but clearly shows that some conditions are better than others with significantly brighter and narrower peaks for the $f_s = 5.5$~GHz in panel b.

\begin{figure*}[!h]
\begin{center}
\includegraphics[scale = 1.0]{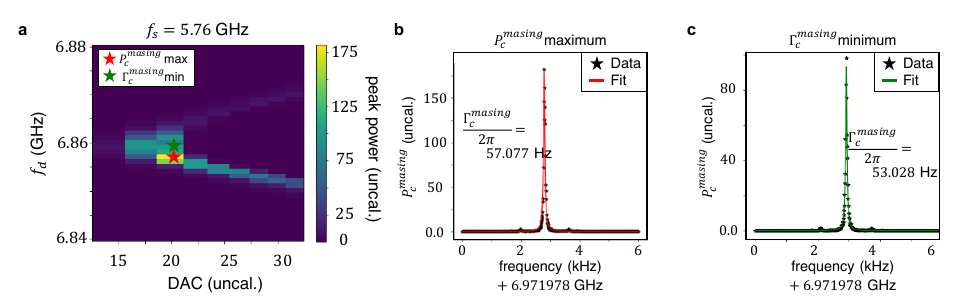}
\caption[\textbf{Masing vs. DAC: fixed $\Phi_{ext}$}]{\textbf{Masing vs. DAC: fixed $\Phi_{ext}$} (a) At a fixed SNAIL bias and transmon bias, we change the DAC ($\propto \Omega$) and thus the drive strength and parametric pumping rate. The DAC values chosen in this measurement are scaling linearly with power which is linear in DAC value squared. The fits from the (b) brightest (c) and narrowest are shown. The minimum bandwidth is $\frac{\Gamma_c^{masing}}{2\pi} = 53.028$~Hz!}
\label{fig:M_masing_vs_pumpfreq_vs_DAC}
\end{center}
\end{figure*}

Another control knob that can be swept is the power of the parametric pump driven to the SNAIL's input port. In Fig.~\ref{fig:M_masing_vs_pumpfreq_vs_DAC} we hold the SNAIL and transmon biases constant and sweep the pump input power (x-axis on the colorplot) and pump frequency (y-axis on the colorplot). We find that the maximum output power and minimum bandwidth are not at the maximum input power. This can indicate that the SNAIL is saturated or that the modes move in frequency due to the high power pump. If the SNAIL and/or transmon shift in frequency, then we may not drive the needed combinations of drive frequencies to hit the needed transitions. 

\begin{figure*}[!h]
\begin{center}
\includegraphics[scale = 1.0]{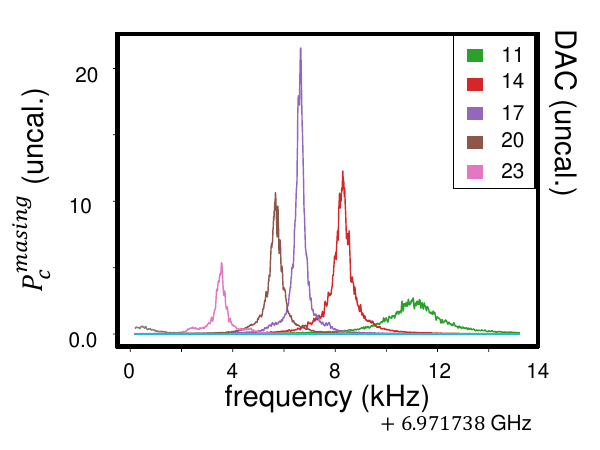}
\caption[\textbf{Masing vs. DAC: fixed $f_d$, $\Phi_{ext}$}]{\textbf{Masing vs. DAC: fixed $f_d$, $\Phi_{ext}$} At a fixed SNAIL bias, transmon bias, and pump frequency, we change the DAC. The maser becomes `better': gets brighter (higher in power output) and narrower (smaller bandwidth in frequency space) for increased DAC until $17$~DAC units, before getting `worse' at higher DAC. This is likely because the pump becomes detuned with increased drive power and photon number. All of these outputs come from the same masing cavity.}
\label{fig:M_masing_fixedbias_changeDAC}
\end{center}
\end{figure*} 
%%data is from 20240210 cooldown%%

This is illustrated by Fig.~\ref{fig:M_masing_fixedbias_changeDAC} which shows individual peaks found at a fixed SNAIL bias, transmon bias, and pump frequency with changing input pump power. There are distinct operating conditions that produce the `best' overall maser, as a function of all of the operating conditions and biases together. This figure illustrates that the maser's stimulated response is impacted by the state of the artificial atom and how it re-populates to feed more coherent light into the cavity.

\begin{figure*}[!h]
\begin{center}
\includegraphics[scale = 1]{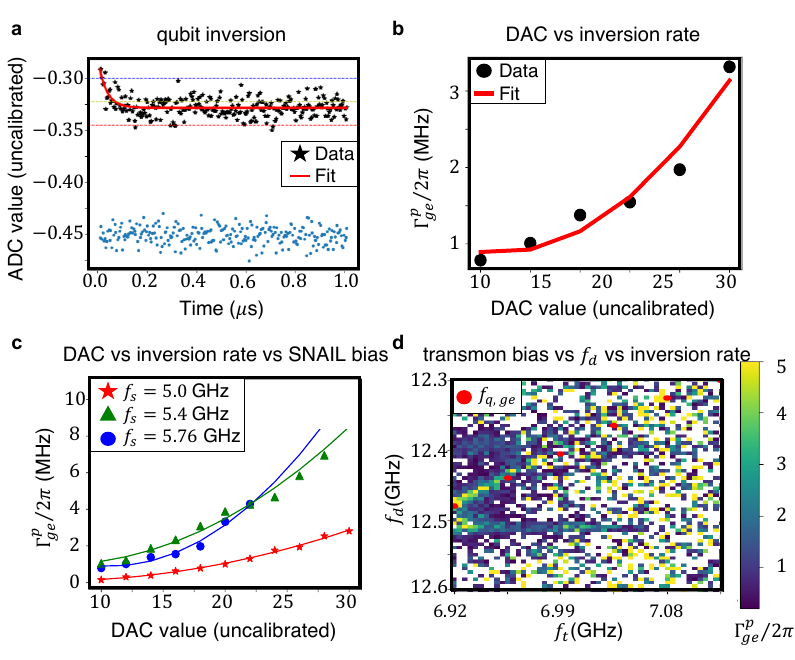}
\caption{\textbf{Qubit inversion} (a) At a fixed $\Phi_{ext}$, $f_d$, pumping strength $\epsilon_p$, we apply the parametric pump and dispersively measure the transmon's state using the qubit
diagnostic readout resonator. We measure  $\Gamma^p_{ge}/2\pi= 4.67$~MHz. (b) At that same $\Phi_{ext}$, $f_d$, we increase the drive strength $\epsilon_p$ through increasing the DAC output on our AWG. This increases $\Gamma^p_{ge}$. The curve follows a quadratic speed up with respect to $\Gamma^p_{ge}$. (c) Repeating the experiment of (b) at different SNAIL bias conditions. Some qubit points are better inverted than others, due to improper detunings in our pump frequency. (d) Sweep transmon bias (x-axis) and parametric drive frequency $f_d$ (y-axis) to find the maximum inversion rates follow the qubit’s frequency trajectory (red circles).}  
\label{fig:qubit inversion}
\end{center}
\end{figure*} 

\begin{figure*}[!h]
\begin{center}
\includegraphics[scale = 1]{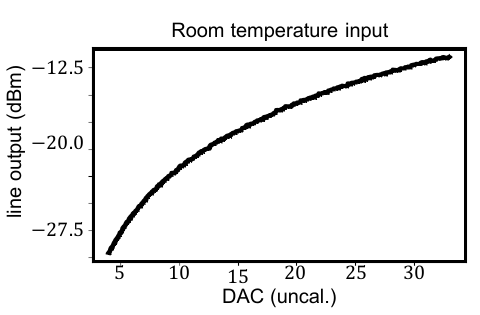}
\caption{\textbf{Room temperature drive strength vs. DAC value} Peak power input of the SNAIL drive at \(12.4\)GHz before going into fridge as shown in Fig.\ref{fig:cryo_diagram} . Plotted as the function of DAC value. }  
\label{fig:drive strength}
\end{center}
\end{figure*} 

\bibliographystyle{apsrev4-1}
\bibliography{refs.bib}